**Low Cost Sensor Networks; How Do We Know the Data are Reliable?**

David E Williams

School of Chemical Sciences and MacDiarmid Institute for Advanced Materials and Nanotechnology

University of Auckland, Private Bag 921019, Auckland 1142, New Zealand

**Abstract**

Plausibility of data from networks of low-cost measurement devices is a growing and important contentious issue. Informal networks of low-cost devices have particularly come to prominence for air quality monitoring. The contentious point is the believability of data without regular on-site calibration since that is a specialist task and the costs very quickly become very much larger than the cost of installation in the first place. This article suggests that approaches to the problem that involve appropriate use of independent information have the potential to resolve the contention. Ideas are illustrated particularly with reference to low-cost sensor networks for air quality measurement.

**Keywords**

**Introduction**

Amongst other benefits, the "Internet of Things" is touted as a ubiquitous measurement storage and communication system, delivering data with high spatial and temporal resolution, from all sorts of different devices. Informal networks of low-cost devices have particularly come to prominence for air quality monitoring [1,2], because, in the face of knowledge that air quality can vary markedly in space and time , they appear to provide data where it is needed and immediately useful: adjacent to residents and not somewhere perceived as remote or inappropriate. Advances in data management, processing, and communications have made it financially and logistically conceivable to operate a spatially dense network of monitors with high temporal resolution. Low-cost devices for this purpose are vigorously and emotively promoted ("what really is in the air around us" : https://airqualityegg.com/home ). Such networks could have the potential to resolve the complex spatial and temporal heterogeneity of air pollution concentrations in urban centres in near-real time. This would make it possible to answer new questions about the underlying causes of poor air quality, ensure more accurate modelling and prediction at local scales [3,4], improve the ability to identify the links between air quality and human health or environmental degradation, identify potential air pollution "hot spots" and enhance the ability to quantify the impacts of pollutant mitigation techniques [5]. Indeed, such developments have been termed a "changing paradigm" for air quality measurement [6]. However, these possibilities can only be fully realised if the measurement uncertainty is low enough to be useful (and what this means needs to be defined) and the maintenance requirement to provide reliable data from large numbers of low-cost sensors is low. The weak point is the measurement technology: the sensors that actually transduce a chemical composition into an observable signal. Any measurement device needs calibrating: the Council of Gas Detection and Environmental Monitoring (http://cogdem.org.uk/newsite/?page=Industrial ) succinctly expresses the view of professional and regulatory organisations: "Although products may be manufactured to national or international standards, the actual measurements taken are only as good as the calibration of the instrument used to do the measuring". For dense networks the cost of calibration and maintenance can very quickly become very much larger than the cost of installation in the first place. The data have to be believable without the need for regular on-site calibration, because the necessary network scale may be very large – hundreds or more of instruments. This is an extremely challenging problem. The following paragraphs suggest a formal approach and then put current efforts into this context.

**Framing the problem**

Since one of the objectives of measurement is to decide whether to take action, a useful philosophical approach is a Bayesian framework, thinking about decisions in the light of uncertainty in information. Here, one asks about the probability of a result, where probability is interpreted as 'degree of belief' in the result, and the relative cost of action and inaction. This is a useful point of view because it enables one to see beyond the matrix algebra and asks a simple question, as framed by de Finetti [7]: what am I prepared to bet on this result being correct? Framed that way, the question immediately focusses attention on the assumptions either explicitly or implicitly made in going from an observation or measurement to a reported result. If the assumptions are hidden, inappropriate, difficult to determine, or unacceptable then the believability of the result will be low. Expressed in other words, Jaynes [8] develops the idea that the expression of probability assigns a degree of plausibility given the evidence, which is revised in the light of new evidence as this becomes available. That approach highlights the evidence. Usually, 'evidence' takes the form of calibration, in which the measurement provided by the instrument is compared to a standard. This process is repeated regularly based on experience with the instrument, such as knowledge of its drift and interference characteristics, known failure modes and their characteristic manifestations, and the reliability of auxiliary apparatus such as sampling pumps or filters. Alternatively, a calibration certificate supplied by the instrument manufacturer may perhaps be relied on. However, for low-cost sensor networks, conventional calibration is too expensive and indeed may be entirely unfeasible, and calibration certificates may not be issued or may have limited duration. Experience of devices in the field which might provide guidance on instrument performance and longevity of calibration may be limited, failure modes not necessarily known, auxiliary apparatus (being itself low-cost) not necessarily reliable, and drift and interference effects ill-characterised and not necessarily small. Thus, although experimental networks have certainly demonstrated feasibility eg [3,4,9-13] and the number of examples is rapidly increasing, based on a conventional view the quality of the data is unknown and the degree of belief in the results would realistically be small. Thus, a conflict arises between the cost of measurement and the required density of instrumentation: a measurement to standards required by regulatory agencies requires expensive, highly-maintained equipment, at carefully chosen sites that obtain reliable samples, yet, as noted by Li et al in a recent study [14], "quantifying pollutant spatial patterns with high fidelity (e.g.,<2 ppb $NO_2$ or<1 µg $m^{-3}$ $PM_1$) seems unattainable in many urban areas unless the sampling network is significantly dense, with more than one or two nodes per $km^2$." One way forward is to think about 'evidence'. For example: 'What constitutes evidence in relation to the results delivered by a sensor network?'; and 'How can evidence be used to increase confidence in the plausibility of the results?' A starting point is some common-sense principles set out in Table 1.

**A philosophical framework based on the theory of measurement**

The theory of measurement and of uncertainty in measurement, as for example expounded by Cox et al. [15] provides a useful and rigorous framework. In practise, it may be computationally too demanding and reliant on significant amounts of historical data, but it does provide a very useful philosophical framework. Three stages are defined: "observation", "restitution", and "measurement". Observation produces an observable output caused by the measurand and thus dependent upon it. Restitution describes the process of estimating the measurand from the observable output. Measurement describes the whole chain, from observation to result. The process involves assumptions, knowledge and models at each stage: without clarity about all of these, the 'believability' of the result cannot be established.

If *y* denotes an observation, viewed as a single instance of a random variable, *Y*, and *x* denotes the particular value of the measurand, again viewed as a single instance of a random variable, *X*, then a model for the measurement instrument describes the probability of observation, *y*, given the instance, *x*. The observation model for the instrument (the probability $\mathbb{P}(y|x,\theta)$ ) in general involves a set of parameters, $\theta$, such as a slope and offset, and coefficients describing any known, measured interferences. The measurement model is the inverse, specifying the probability : $\mathbb{P}(x|y,\theta)$. Restitution is the process of determining the probability of instance, *x*, given observation, *y*, and other knowledge, information or models, $\mathcal{I}$ : $\mathbb{P}(x|y,\theta,\mathcal{I})$. Implied in the statement of a probability is the assumption of a model for the distribution, so restitution describes the distribution to be assigned to the measurand on the basis of the observation, *y*, and the assumptions, models and other information applied to the measurement instrument and the process being measured: for example, information pertinent to knowledge of the parameters $\theta$. If the parameters are determined from the information then:

$$\mathbb{P}(x|y) = \int \mathbb{P}(x|y,\theta)\mathbb{P}(\theta)\mathrm{d}\theta = \int \mathbb{P}(x|y,\theta)\, \mathbb{P}(\theta|\mathcal{I})\, \mathbb{P}(\mathcal{I})\, \mathrm{d}\mathcal{I} \qquad (1)$$

Measurement is the reporting of the probability density function (pdf) of the estimate, $\hat{x}$, given the estimated distribution of *x*: $\mathbb{P}(\hat{x}|x)$. For environmental sensors, the reported value may be a mean evaluated over some time, *t*, $\mu_{x,t}$ , and the user may be interested in the probability that $\mu_{x,t}$ lies within some range or whether $\mu_{x,t}$ exceeds some threshold, $\mu^*$. As formulated in the theory of measurement, the procedure applies to a single instrument. For instruments in a network, the ideas would need to be developed both for the individual measurement points and for the pattern revealed by the network either as a whole, or within a suitably limited locality. How this might be done is discussed later.

**What is "calibration" ?**

Calibration is the conventional means by which the information to determine the parameters, $\theta$, is obtained [16,17]: "calibration is the complex of operations aimed at obtaining the conditional distribution that characterises the observation " [17] . This is the probability distribution to be assigned to the (unknown) 'true' value, *x*, given the observation, *y*, and the instrument parameters, $\theta$. For sensors used within a network, equation 1 indicates a way towards resolving the 'plausibility' problem, which is that in a network and given knowledge about the particular problem domain, there may be significant *independent* information that can be used to determine the conditional distribution of the parameters. An issue is that the computational overhead required to implement such a procedure could be very large and so not practical for a network aiming a high spatial and temporal resolution. Furthermore, as Weise [18] points out: "the result of Bayesian inference is uniquely determined once the distribution has been chosen"; and this is the crunch: what distribution function should be chosen that reasonably represents the problem domain?  This is where important assumptions may be hidden.  Nevertheless, the formal framework helps to clarify how one might go about establishing some degree of plausibility for network results, and why some methods might be preferred over others.

**"Independent information": issues with proprietary models embedded in sensors or networks.**

What might be the independent information? Examples include:

a) Knowledge of the characteristics of the sensor, embodied in an observation model for the response, $y = f(x, \theta)$. A simple example is the linear model for the response: $y = a_0 + a_1 x$.
b) Prior expectations on the parameters $\theta$, based on measurements performed some time previously, or on general knowledge of the sensor characteristics derived for example from a long-term body of observations.
c) Known influences on the sensor.
d) Known general characteristics of the measurement problem such as the general variation of the characteristics of the unknown $x$ over a geographic region, expected diurnal variations associated with motor vehicles or sunlight, variations associated with different land-use (eg proximity of roads, industry or open country, high-rise buildings, waterways).
e) Measurements with other techniques, such as reference instruments at particular locations, satellite observations, long-term average sample measurements.
f) Correlations with meteorology and expected spatial patterns as a result – rainfall, wind direction and speed.
g) Computational models, e.g. dispersion models; land-use regression; interpolation models based on sparse reference data or on more densely distributed time-averaged measurements made in specific campaigns using for example sample collection, sampling filters or Palmes diffusion tubes.

In principle, if the influences on the sensor are known, their pdf measured with known uncertainty and the relationship to the sensor output clear, then the formal theory of sensor data fusion leads to the pdf of the measurement result given the observations [19]. For low-cost sensor networks, however, the information required may not be generally available and the computational burden may be too high. If the models are not clear, then assessment of plausibility cannot be made. One particular domain of contention is the application of machine learning methods (particularly in respect of a) – c) above) and of proprietary algorithms (empirical; not based on any underlying physical law) for deriving the reported result from raw sensor data. The issue is that what is reported is not a measurement result from an instrument but the result of a (hidden) model for the measurand : in the notation above, the estimate $\hat{x}$ without any information about the raw data, the parameters or the conditional distributions, and hence no knowledge of the uncertainty, other than (perhaps) a performance measure on a test set and (perhaps) little information on the transferability of the training set used to derive the model. If an independent assessment of plausibility cannot be made, then clearly there will be contention over the meaning of the results. Concern about the plausibility of machine learning approaches, particularly the interpretability of models, is not confined to the study of chemical measurement [20].

**Discussion of sensor issues in relation to air quality measurement**

The issues can be illustrated with reference to current work on the development of networks of low-cost instruments for air quality measurement: specifically particle measurement as cumulative mass under than a defined size cutoff; and pollutant gas concentration measurement (ozone and nitrogen dioxide). For both of these, threshold exposures averaged over different periods of time are defined in legislation.

Particle measurement is tricky [21]. Results are reported as mass per unit volume assuming spherical particles, and the thresholds are given in the same units. Immediately, there is an issue: particles are not in general spherical and unless cumulative mass is directly measured a particle density or other parameter to convert the measurement into mass needs to be assumed. However, even reference instruments use a variety of methods other than mass: β-particle attenuation, and light scattering. . A cyclone or impactor may be used as an inlet size filter the cut-off radius of which is an aerodynamic radius determined with an assumption concerning the particle density and shape. The light scattering instruments [21] may use extinction along a sufficiently long light path, or single particle counting, where an air stream is drawn through a laser beam and scattered light pulses are binned according to pulse height. Light scattering is a complex function of particle size, shape and refractive index. An approximate average scattering curve is used and a particle size distribution function may be assumed as part of the data fitting procedure. Since particles can be hygroscopic, changing size significantly with water vapour pressure, or they may absorb volatile components during emission at source, heated inlets may be employed. Given the assumptions, even with reference-grade instruments correlation between instruments using different measurement methods can be less than ideal, and dependent on the particle type (source) and atmospheric conditions (temperature, humidity). Finally, the bulk of particles by number is at very small sizes, though these may not contribute significantly to the distribution by mass: with optical light scattering as the measurement method, very small particles need high-grade instruments for their detection. Given that, it is unsurprising that low-cost devices can have poor correlation with reference methods [22-26]. These devices can indeed have excellent correlation between themselves, indicating that the measurement and algorithm is consistent from one device to another of the same manufacturer (they are precisely inaccurate). They do show signals that rise and fall in general correlation with reference methods, and this can be useful. However, the method by which signal, from the photodetector, is converted to particle count, and hence the assumptions that are made, is not visible, and the fact that different manufacturers make different assumptions is illustrated by the poor correlation between results from devices of different manufacturers [25]. 'Plausibility', that a linear model relates reported count to reference count and is valid under specified conditions of water vapour pressure, temperature and source attribution, is established by long co-location with reference analysers in the environment. Such co-locations establish the probability distribution function associated with the measurement model, in relation to the chosen reference and under the defined conditions of the comparison, and of course assume that the measurement model or any critical component have not been changed by the manufacturer. Plausibility of the reported result as opposed to plausibility of the measurement model requires other information, as may be derived from d) to g) above, for example. Since the general objective of measurement is comparison of the results with a (legally defined) threshold, and action to be taken as a consequence, the further step, of data fusion with additional, independent information in order to develop an assessment of the reliability of the reported result, would seem to be essential.

Electrochemical cells applied to measurement of gaseous pollutants illustrate different aspects of the general problem of plausibility of reported results, and hence whether the information delivered is useful in forming decisions [27]. These, like the low-cost light-scattering devices, are conceptually simple but surprisingly complex in practise [28,29]. They were originally developed for health and safety applications – hence measurement at much higher concentrations than those relevant to air quality measurement – and are extremely effective and well-established for that purpose. However, in relation to pollution measurement, there is relatively little reported raw data showing the electrochemical signal (the current through the cell) with resolution appropriate for environmental measurement. What there is, shows that the current measured in a laboratory environment is very stable, but in the outside atmospheric environment is very noisy [3]. The noise-averaging procedure is a hidden assumption for these devices; the way that the noise may vary as a consequence of changes in water vapour pressure or temperature has not been reported. The devices also have an offset current that is large in comparison with the signal current, hence small variations in offset current, such as might easily be caused by the movement of the meniscus at the 3-phase gas-electrode-electrolyte interface, have a significant effect. Without some unambiguous estimation of the 'zero' at sufficiently short intervals to account for atmospheric variation the plausibility of the result must be in question. The devices are also particularly sensitive to rapid changes in atmospheric humidity (most likely due to meniscus fluctuations) and have some significant cross-interferences – ozone with nitrogen dioxide is a particular, important example [3]. Machine-learning methods have been adopted to manipulate the input (the current in the electrochemical cell) to deliver the output (a reported concentration) given auxiliary measurements, some of which use other electrochemical cells with similar issues, or low-cost devices for measurement of humidity and temperature. The resultant algorithms may be embedded in commercial systems. What this means for assessment of plausibility, is that the parameters, $\theta$, incorporated into the measurement model may be unknown and their probability distribution function cannot be evaluated. Again, data fusion with additional, independent information in order to develop an assessment of the reliability of the reported result, would seem to be essential [29].

Sensors based on semiconducting oxides as gas-sensitive resistors have been much investigated for application in environmental measurement. Their characteristics have been much-studied and are well-understood [30,31]. As for electrochemical cells, the issue of cross-interference and particularly the effect of water vapour pressure is important. There is one particular example where such sensors have proved excellent: for environmental measurement of ozone [13,32-34]. The key is the extremely large resistance change induced in heated $WO_3$ by trace ozone, much larger than the effect of any other trace gas in the environment with the possible exception of $H_2S$, and the use of flow and temperature oscillation to cancel the effects of water vapour and continually reset the sensor and determine the zero [32,35,36]. This method represents a significant improvement over the more usual method of operation at constant temperature with auxiliary measurement and compensation for the effects of variable water vapour pressure [37]. Hence with the use of temperature and flow modulation, the measurement model is robust. The issue is that dirt deposits over time onto the sensor element and onto the inlet filters, leading to a decrease in the gas flow rate across the sensor, hence to ozone decomposition in the inlet system and hence to a decrease in the slope of the response ($a_1$, above). Therefore, independent information is needed to detect and correct this drift. Such information is available if the sensors are used within a hierarchical network that includes well-maintained and rigorously calibrated reference instruments [34,38,39].

**Incorporation of independent information into sensor correction algorithms**

Potentially powerful approaches to the issue of plausibility use independent knowledge about the problem domain. A number of methods have been proposed to incorporate independent information such as in d) – g) above into correction algorithms for 'blind' or 'semi-blind' calibration of sensors in a network. The most important aspect to understand of these works is whether the assumptions are valid for the problem domain or instruments employed: that is, the assumptions and restrictions implicit in the statement of $\mathbb{P}(\theta)$ in eq. 1. Table 2 lists some examples from the literature. As Table 2 shows, the assumptions can be very restrictive, and may be difficult to verify in practise: verification may, indeed, involve significant labour hence vitiating the presumed cost advantage of a 'low-cost' network.

Additionally, there are methods which aim to fuse instrument data, land-use, satellite data, and qualitative information into models for interpolating between both sparsely distributed measurement sites [40-42] and between low-cost sensor sites [3,43]. The philosophical approach behind these ideas [41,44,45] is similar to that leading to eq 1. Thus, in principle the sensor measurement model can be included in the computation with the sensor parameters as variables. A key assumption is how correlation between neighbouring sites decays with distance (the 'semivariogram' model) as well as the model for the pdf of the environmental variables [45]. The approach is similar to methods used in image restoration, which proceed by defining neighbours of a given point, then adjust 'colour' (which here would be interpreted as the sensor parameters) in order to be consistent with a model for the spatial correlation within the neighbourhood [46,47]. Similar ideas have been used to combine computational models for pollutant emission and dispersion with uncertain sensor data in order to derive interpolated maps across a region [48]. An issue of course is that, as the model complexity increases, the computational burden rises and the transparency of assumptions decreases. Thus, there is a need to be very clear about the measurement model, to demonstrate unequivocally that the measurement model is reliable, and to show how the derived sensor parameters vary over time [39]: a smooth temporal variation of derived parameters rather than large random variations would indicate 'plausibility'.

**Plausibility and usefulness**

Ultimately, whether sensor network data are plausible is connected to the question of whether they are useful. Thus, if the network data reveal patterns which seem plausible based on other experience, or which on investigation have an explanation, then the data themselves can be accepted because they have revealed something useful. For urban air quality measurement, land-use regression models for mean concentrations evaluated over a few weeks can be developed on small spatial scales, using Palmes diffusion tubes or hand-held sensors [49-52] or by using the sensor data themselves. If enough sensors sample environments with similar values for the covariates, then (as in a standard analysis of variance) comparison of the deviations from the general regression of individual sites with similar or different covariates can say something about the contribution of sensor variance to the overall variance. In a sense, this is treating sensors with similar values of the covariates as 'neighbours'. On the 'microscale' the significance of particular urban design features can be evaluated. These models are interesting and useful because they reveal urban design features that are of importance for personal exposure. Now, the hourly mean of low-cost sensor measurements can be correlated with the LUR-modelled mean: a very simple linear model can be applied [3]. With this rather simple data fusion, the microscale land-use model and its overall correlation with sensor data estimates the average variation, with high spatial and temporal resolution, and can be used to indicate locations where personal exposure may be highest – pedestrian crossings at intersections or under awnings at bus stops are obvious examples that such analysis reveals [3]. The computational burden is very small. The deviations of each individual sensor result from this correlation can then be examined: are the deviations random, or systematic, or do they show a consistent temporal variation? If the latter, then specific urban design features of the sensor location can be examined to see if there is a plausible explanation – for example, the effect of parked cars impeding diesel buses leaving a bus stop [3]. In that case, the sensor data are both plausible and useful. Given that air quality, and its management by measures that impact on personal freedoms, is a highly conflicted societal issue, objective evaluation of the consequences of control measures using high-density measurement networks in conjunction with the methods described above would seem to be an important development. Examples could include evaluation of the effect of local urban design changes such as vegetation or barriers [5,53-56] or changes to car parking or traffic light sequencing, as well as evaluation of major initiatives such as low-emission zones or location- and time-based emission charging. A rigorous approach to the question of plausibility of data will, however, be essential if such developments are to be realised. The connected use as outlined above of low-cost sensor networks, reference networks, models and supplementary data such as satellite observations seems a way to resolve the conflicts about plausibility and application, but this will require a significant effort and cannot be achieved without transparency through to the raw data and the restitution models, for which a conflict with the proprietary interests of instrument manufacturers will have to be resolved.

**Acknowledgement**

This work has been supported by the NZ Ministry for Business, Innovation and Employment, contract UOAX1413. The author acknowledges the award of a Fellowship of the Institute of Advanced Study, Durham University, UK, and many enlightening discussions with colleagues in environmental science and statistics.

Table 1

Sensor network elements

- "sensor"
    - The assembly of indicating element and its control electronics, sampling system, raw data measurement, data conversion and transmission.
        - Otherwise known as 'low-cost instrument' with the indicating element called the 'sensor'. In this context it is the entire assembly, that delivers data to the network system.
    - **moral**: know how your sensor behaves in the environment, in the long-term, and know its foibles
- "network"
    - The sensor locations and their connected arrangement
    - **moral**: understand its purpose, and the constraints on location
- "system"
    - Maintenance, calibration, correction and interpolation
    - Delivers the result to the user
    - **moral**: understand the user requirements
        understand what is meant by "reliability" in relation to the network purpose
        understand the problem domain and how its characteristics can be used to give confidence in data reliability

Table 2 Example methods for 'blind' or 'semiblind' assessment of sensor calibration in a network

| method | assumptions | Ref. |
|---|---|---|
| Linear measurement model for sensors. Identify principal component vector for the spatial variation of the measurand (spatial variation can be expressed as a linear function of a set of basis functions of the sensor locations). Compute slope for individual sensors to minimise spatial variation orthogonal to the principal component vector. Alternatively framed as a 'consensus' problem | Linear measurement model. Spatially-smooth variation of measurand for all times of measurement. Measurement space is spatially over-sampled: ie spatial frequency of measurement points is significantly greater than lowest spatial frequency associated with the measurand variation. To compute offset requires at least one "true" measurement location<br><br>Alternatively, for the unknown offset, additionally assume that immediately on deployment, sensors are reliable (reliable initial calibration; no unknown drift). Or, at a certain time instant at most one sensor would have an unknown drift. Assume smooth parameter drift develops a Gaussian distribution of parameters. | 57-59<br><br><br><br><br>60-62 |
| Random rendezvous. Mobile sensors have single point calibration when they pass a reference station, and may transfer this calibration to other (fixed) devices as they pass. Slope and offset derived from multiple encounters between mobile sensor and reference stations. | Linear measurement model. Measurement drift assumed slow with respect to time required to acquire sufficient calibration points. Weighting function applied according to time since last encounter with a reference station. Alternatively, temporal variation of the field being sensed is slow with respect to time to accumulate sufficient observations. | 63-67 |
| Random rendezvous, with sensors forming groups that rearrange over time, with specific assumptions about the sensors | (a) Mean value of all sensor offset drifts is zero, with a sufficient number of sensors.<br>(b) Gain and offset are (truncated) Gaussians with known mean and variance. Sensors in clusters observe the same field, which has a known dynamic behaviour (relationship of signal at time $t$ to signal at the previous observation time) | 68<br><br>69 |
| Moment matching | Sensor devices moving in the same region hence signal statistics over time are the same; signal field statistics stable in time. Expected value of gain and offset for ensemble of sensors is known | 70 |
| Chain co-location: co-locate sensor with reference site; move to buddy with test site for a period; move original test site sensor to buddy at next test site etc eventually back to reference station | Sensor parameter variation is slow with respect to time scale for the sensor progression around the sites. | 71 |
| Hierarchical network, of a number of well-maintained reference stations together with a much larger number of low-cost sensors. For each sensor site, identify a 'proxy' that has a similar probability distribution of measurement values, evaluated over a suitable period. Determine sensor parameters by matching sensor data distribution to the proxy. For a 2-parameter (slope and offset) problem, match distributions by matching mean and variance. For a multi-parameter linear problem, given auxiliary measurements, adjust parameters to minimise Kullback-Leibler divergence between sensor and proxy distributions | Requires assessment of proxies using data from the reference network. Land-use criteria used. Reference network needs to be extensive enough to sample different land use, with high data availability.<br>The information is used to select the proxy site (index, $m$, below) then the proxy data, $[x_m]$ are used to estimate the sensor parameters. Formally:<br>$\mathbb{P}(x_0|y_0) = \int \mathbb{P}(x_0|y_0, \theta)\mathbb{P}(\theta|[x_m])\mathbb{P}(m|\mathcal{I})\mathbb{P}(\mathcal{I})d\mathcal{I}$. Practically, the reliability is estimated by co-locating sensors with reference stations and comparing the derived sensor result with the reference. A key assumption is that reference sites are representative of the area being monitored. | 34,39,49,52 |
| Hierarchical network. For each sensor site estimate mean and variance over a suitable period by spatial interpolation using the calibrated sites. Compute sensor slope and offset from estimated mean and variance. | Assumptions implicit in kriging as the interpolation method: smooth spatial variation of mean and variance of concentration field; Gaussian semivariogram, spatially uniform. Requires a sufficient spatial density of calibrated stations for reliable interpolation estimate. | 37 |